\documentclass[doublespacing]{elsart}

\usepackage{icarus}

\usepackage{natbib}

\bibpunct{(}{)}{;}{a}{,}{,}

\usepackage{graphicx}
\usepackage{lscape}

 \usepackage{amssymb}

\newcommand{\BibTeX}{ \textrm{B\kern-.05em\textsc{i\kern-.025em b}\kern-.08em
    T\kern-.1667em\lower.7ex\hbox{E}\kern-.125emX} }

\newcounter{ionctr}
\ifx \undefined \ion \newcommand{\ion}[2]{\setcounter{ionctr}{#2}{#1$\;${\small\rmfamily\Roman{ionctr}}\relax}} \fi

\newcommand\arcsec{\mbox{$^{\prime\prime}$}}
\usepackage{color,soul}
\usepackage{ulem}

\begin{document}

\begin{frontmatter}



\title{Evolution of H$_2$O Production in Comet C/2012 S1 (ISON) as Inferred from Forbidden Oxygen and OH Emission\thanksref{McD}}

\thanks[McD]{This paper includes data taken at The McDonald Observatory of The University of Texas at Austin, as well as data collected at the W.M. Keck Observatory, Maunakea, HI, USA, operated as a scientific partnership among Caltech, UCLA, and NASA, and made possible by the generous financial support of the W.M. Keck Foundation.}


\author[label1]{Adam J. McKay},
\author[label2]{Anita L. Cochran},
\author[label3,label4]{Michael A. DiSanti},
\author[label5]{Neil Dello Russo},
\author[label5]{Harold Weaver},
\author[label5]{Ronald J. Vervack Jr.},
\author[label6]{Walter M. Harris},
\author[label7]{Hideyo Kawakita}

\address[label1]{NASA GSFC/USRA, 8800 Greenbelt Rd, Greenbelt, MD 20771 (U.S.A.); adam.mckay@nasa.gov}
\address[label2]{University of Texas Austin/McDonald Observatory, 2512 Speedway, Stop C1402 , Austin, TX 78712, (U.S.A); anita@astro.as.utexas.edu}
\address[label3]{NASA Goddard Center for Astrobiology, NASA GSFC, Mail Stop 690, Greenbelt, MD 20771 (U.S.A.); Michael.A.Disanti@nasa.gov}
\address[label4]{Solar System Exploration Division, Mail Stop 690, Greenbelt, MD 20771 (U.S.A)}
\address[label5]{Johns Hopkins University Applied Physics Laboratory, 11100 Johns Hopkins Rd., Laurel, MD, 20723 (U.S.A.); neil.dello.russo@jhuapl.edu, Hal.Weaver@jhuapl.edu, Ron.Vervack@jhuapl.edu}
\address[label6]{Lunar and Planetary Laboratory, University of Arizona, 1629 E University Blvd., Tucson, AZ, 85721 (U.S.A);wharris@lpl.arizona.edu}
\address[label7]{Koyama Astronomical Observatory, Kyoto Sangyo University, Motoyama, Kamigamo, Kita-ku, Kyoto 603-8555, Japan; kawakthd@cc.kyoto-su.ac.jp}
\begin{center}
\scriptsize
Copyright \copyright\ 2018 Adam J. McKay, Anita L. Cochran, Michael A. DiSanti, Neil Dello Russo, Harold Weaver, Ronald J. Vervack, Walter M. Harris, Hideyo Kawakita
\end{center}


%
%
%
%
%


\end{frontmatter}



\begin{flushleft}
\vspace{1cm}
Number of pages: \pageref{lastpage} \\
Number of tables: \ref{lasttable}\\
Number of figures: \ref{lastfig}\\
\end{flushleft}


\begin{pagetwo}{H$_2$O Production in Comet C/2012 S1 (ISON)}

Adam J. McKay \\
NASA GSFC/USRA\\
8800 Greenbelt Rd\\
Greenbelt, MD 20771, USA. \\
\\
Email: adam.mckay@nasa.gov\\
Phone: 301-614-5701\\

\end{pagetwo}

\begin{abstract}
We present H$_2$O production rates for comet C/2012 S1 (ISON) derived from observations of [\ion{O}{1}] and OH emission during its inbound leg, covering a heliocentric distance range of 1.8-0.44 AU.  Our production rates are in agreement with previous measurements using a variety of instruments and techniques and with data from the various observatories greatly differing in their projected fields of view. The consistent results across all data suggest the absence of an extended source of H$_2$O production, for example sublimation of icy grains in the coma, or a source with spatial extent confined to the dimensions of the smallest projected field of view (in this case $<$ 1,000 km).  We find that ISON had an active area of around 10 km$^2$ for heliocentric distances R$_h$ $>$ 1.2 AU, which then decreased to about half this value from R$_h$=1.2-0.9 AU.  This was followed by a rapid increase in active area at about R$_h$=0.6 AU, corresponding to the first of three major outbursts ISON experienced inside of 1 AU.  The combination of a detected outburst in the light curve and rapid increase in active area likely indicates a major nucleus fragmentation event.  The 5-10 km$^2$ active area observed outside of R$_h$=0.6 AU is consistent with a 50-100\% active fraction for the nucleus, larger than typically observed for cometary nuclei.  Although the absolute value of the active area is somewhat dependent on the thermal model employed, the changes in active area observed are consistent among models. The conclusion of a 50-100+\% active fraction is robust for realistic thermal models of the nucleus.  However the possibility of a contribution of a spatially unresolved distribution of icy grains cannot be discounted.  As our [OI]-derived H$_2$O production rates are consistent with values derived using other methods, we conclude that the contribution of O$_2$ photodissociation to the observed [\ion{O}{1}] emission is at most 5-10\% that of the contribution of H$_2$O for ISON.  This is consistent with the expected contribution of O$_2$ photodissociation if O$_2$/H$_2$O $\sim$ 4\%, meaning [\ion{O}{1}] emission can still be utilized as a reliable proxy for H$_2$O production in comets as long as O$_2$/H$_2$O $\lesssim$ 4\%, similar to the abundance measured by the ROSINA instrument on Rosetta at comet 67P/Churyumov-Gerasimenko. 

\end{abstract}



\section{Introduction}

\indent Comet C/2012 S1 (ISON) was a dynamically new comet from the Oort Cloud, meaning it was making its first passage through the inner Solar System~\citep{Novski2012}.  It passed within 3 solar radii of the Sun in late November 2013, making ISON a sungrazing comet.  Many sungrazing comets have been discovered, most belonging to a dynamical family called the Kreutz group.  However, ISON was unique in two respects: 1) it was dynamically new, whereas the Kreutz group comets are not, and 2) more importantly, ISON was discovered over a year before its perihelion passage, meaning observations could be planned well in advance to follow its plunge toward the Sun.  This meant that ISON could be observed over a very large range of heliocentric distances, which was not possible for previous sun-grazing comets, potentially providing new insights into cometary composition and activity.  Because of this unique opportunity, a world-wide observing campaign was organized, providing an unprecedented data set.\\

\indent Comet ISON experienced very irregular activity levels as it approached the Sun, experiencing several outbursts that may or may not have been correlated with fragmentation or disruption events~\citep[e.g.][]{SekaninaKracht2014}.  Measuring the H$_2$O production rate throughout its inbound trajectory provides a window into the sublimation processes driving the outbursts.

While direct observations of H$_2$O are possible at IR wavelengths~\citep[e.g.][]{DelloRusso2011}, H$_2$O production can also be measured indirectly through its dissociation products H, O, and OH.  The main emission mechanism for the [\ion{O}{1}]6300~\AA~line is prompt emission after photodissociation of an H$_2$O molecule~\citep{FestouFeldman1981}.  Therefore observations of this line have been used as a reliable proxy for H$_2$O production in the past~\citep[e.g.][]{Morgenthaler2001, Morgenthaler2007, Fink2009, McKay2015}.  Also, OH in the coma results from photodissociation of H$_2$O, and OH is often used as a proxy for H$_2$O at NUV~\citep[e.g.][]{AHearn1995, Opitom2015}, IR~\citep[e.g.][]{Bonev2006, DelloRusso2011}, and radio wavelengths~\citep[e.g.][]{Biver2002}.  Lyman-$\alpha$ emission has also been used to derive H$_2$O production rates, particularly with the SOHO satellite~\citep[e.g.][]{Combi2013}. 

\indent We present analysis of observations of [\ion{O}{1}] and OH emission in comet ISON as a proxy for H$_2$O production.  We obtained multiple observations spanning the time period from September-November 2013, covering a heliocentric distance range of 1.8-0.44 AU.  Our observations of [\ion{O}{1}] in particular, which have excellent sensitivity due to the abilities of optical detectors and the prompt nature of the [\ion{O}{1}] emission mechanism, cover larger heliocentric distances than other studies, providing a characterization of ISON's activity earlier in the apparition.  This paper is organized as follows: in section 2 we describe our observations and reduction and analysis procedures.  Section 3 presents our results.  In section 4 we discuss our results in the context of the wider observing campaign, including comparison to other measurements of the H$_2$O production rate obtained with different methods.  Section 5 presents a summary of our conclusions.

\section{Observations and Data Analysis}
\subsection{Observations}
\indent We obtained spectra of C/2012 S1 (ISON) using three facilities. The ARCES instrument is mounted on the Astrophysical Research Consortium 3.5-m telescope at Apache Point Observatory (APO) in Sunspot, New Mexico.  ARCES provides a spectral resolution of R $\equiv$ $\frac{\lambda}{\Delta\lambda}$ = 31,500 and a spectral range of 3500-10,000~\AA~with no interorder gaps.  More specifics for this instrument are discussed elsewhere~\citep{Wang2003}.    In mid-October we obtained spectra using the Tull Coude Spectrograph mounted on the 2.7-m Harlan J. Smith telescope at McDonald Observatory.  The spectral range is the same as ARCES, but the Tull Coude provides twice the spectral resolution, with R $\sim$ 60,000.  However, unlike ARCES, the Tull Coude spectrograph has interorder gaps redward of 5700~\AA.  To account for this, we used a spectrograph setup such that the interorder gaps did not fall on emission features of interest, which for the work presented here constitutes the [\ion{O}{1}]6300~\AA~line.  We obtained observations with the HIRES instrument on Keck I in late October.  We utilized the HIRESb mode, which covers wavelengths from 3000-5700~\AA~at R $\sim$ 48,000 with small gaps in coverage at the edges of the CCD detectors.  Although this mode does not cover the redder wavelengths sampled by the ARCES and Tull Coude data (and therefore not the [\ion{O}{1}]6300~\AA~line), the additional extension blueward into the UV provides observations of the OH A-X (0-0) band at 3080~\AA~that is not sampled in the ARCES or Tull Coude spectra.\\

\indent The observation dates and geometries, as well as standard stars used and observing conditions are described in Table~\ref{observations}.  For all observations we centered the slit on the optocenter of the comet.  All the instruments employed have relatively narrow and short slits, specifically 1.6\arcsec$\times$3.2\arcsec (ARCES), 0.86\arcsec$\times$7\arcsec (HIRES), and 1.2\arcsec$\times$8.2\arcsec (Tull Coude).  We used an ephemeris generated from JPL Horizons for non-sidereal tracking of the optocenter.  For short time-scale tracking, the guiding software uses a boresight technique, which utilizes optocenter flux that falls outside the slit to keep the slit on the optocenter.  For the ARCES and HIRES observations, we observed a G2V star in order to remove the underlying solar continuum and Fraunhofer absorption lines.  For the McDonald observations, we obtained observations through a solar port which feeds sunlight directly into the spectrograph.  We obtained spectra of a fast rotating (vsin(i) $>$ 150 km s$^{-1}$), O, B, or A star to account for telluric features and spectra of a flux standard to establish absolute intensities of cometary emission lines.  The calibration stars used for each observation date are given in Table~\ref{observations}.  We obtained spectra of a quartz lamp for flat fielding and acquired spectra of a ThAr lamp for wavelength calibration.\\

\subsection{Data Reduction}

\indent Spectra were extracted and calibrated using IRAF scripts that perform bias subtraction, cosmic ray removal, flat fielding, and wavelength calibration.  We removed telluric absorption features, removed the reflected solar continuum from the dust coma, and flux calibrated the spectra employing our standard star observations.  We assumed an exponential extinction law and extinction coefficients for the observatory site when flux calibrating the cometary spectra~\citep[e.g.][]{Hogg2001}.  More details of this procedure can be found in~\cite{McKay2012} and~\cite{CochranCochran2002}. We determined slit losses for the flux standard star observations by performing aperture photometry on the slit viewer images as described in~\cite{McKay2014}.  Slit losses introduce a systematic error in the flux calibration of $\sim$ 10\%.  Slit viewer images are not available for the McDonald and Keck data sets, so for these data sets we adopted a slit loss value based on the measured seeing and  assumption of a Gaussian PSF.  Accounting for possible variability in seeing and the ideal nature of this assumption, we estimate the uncertainties in slit losses are $\sim$ 20\% in these cases.  For UT September 23 and UT November 20, flux standards taken from UT October 3 and UT November 15, respectively, were used to flux calibrate the spectra since a flux standard star was not obtained those nights.  This may introduce additional uncertainty that is not accounted for in our quoted measurements.\\

\indent With the high airmass ($>$ 2-3) of some of our observations and the slits employed (widths of 0.9-1.6\arcsec, lengths of 3.2-8.2\arcsec), the effect of differential refraction on fluxes obtained at different wavelengths could be significant.  This could be particularly important for OH, whose primary emission band is in the near UV (3100~\AA) and therefore suffers from more significant differential refraction than redder spectral features.  To evaluate the effect of differential refraction, we employ the computed differential refraction as a function of wavelength and airmass from~\cite{Filippenko1982}, and interpolate to the relevant wavelengths and airmasses for our observations.  We calculate the amount of differential refraction and use a Haser model profile convolved with a Gaussian point spread function to estimate the slit loss due to the differential refraction.  For all our observations we estimate the effect of differential refraction is less than 10\% and is insignificant compared to the uncertainties in flux calibration noted above (10-20\%).  The uncertainties quoted in this paper are dominated by the uncertainty in flux calibration and not the signal-to-noise ratio of the data nor differential refraction effects.\\

\subsection{Analysis of [\ion{O}{1}] Emission}
\indent The [\ion{O}{1}]6300~\AA~line is among the emission features present in the ARCES and Tull Coude bandpasses and can be employed to derive the H$_2$O production rate.  The [\ion{O}{1}]6300~\AA~line is also present as a telluric emission feature, and so a combination of high spectral resolution and adequate geocentric velocity (and therefore Doppler shift) is needed to resolve the cometary line from the telluric feature.  For all observation dates except November 20 the cometary and telluric lines are fully separated.  On November 20 the comet was so bright and active that the cometary line is expected to dwarf the telluric feature, therefore any influence of the telluric feature on the measured flux is negligible.  An example spectrum from November 15 is shown in the top panel of Fig.~\ref{spectra}, with a more detailed view of the [\ion{O}{1}]6300~\AA~and [\ion{O}{1}]6364~\AA~lines shown in Fig.~\ref{zoom}.  To measure the line flux we fit a Gaussian function to the line profile.  More details for our line fitting procedure can be found in~\cite{McKay2012}.\\ 

\indent For all of our observations, to check our calibration we measured the flux ratio of the [\ion{O}{1}]6300~\AA~line to the [\ion{O}{1}]6364~\AA~line (hereafter the red line ratio).  As these lines both originate from the $^1$D state, the red line ratio is determined directly by the branching ratio into the different transitions and is not dependent on coma chemistry or physics.  This value is well established at a value of 3.0~\citep{CochranCochran2001, SharpeeSlanger2006}.  However, our observations in September and October yield significantly lower values for the red line ratio, in the range 1.5-2.5.  Examination of the spectra revealed that at the very large Doppler shift that ISON had at the time of these observations (-50 km s$^{-1}$), the [\ion{O}{1}]6300~\AA~line falls precisely on an O$_2$ telluric absorption feature, as shown in Fig.~\ref{raw}.  Although we remove telluric signatures during our reduction process, the removal may not be perfect.  On nights for which we had multiple observations, [\ion{O}{1}]6300~\AA~line intensities have a large amount of scatter, evidence that our [\ion{O}{1}]6300~\AA~line intensities were affected by imperfect removal of this telluric absorption feature from spectrum to spectrum.  There are no telluric or solar absorption features near the [\ion{O}{1}]6364~\AA~line, suggesting this line should not be affected by the presence of these features (see Fig.~\ref{raw}).  The [\ion{O}{1}]6364~\AA~line intensities exhibit significantly less scatter, consistent with the stochastic uncertainties and supporting this conclusion.  Therefore for the September and October data we use the  [\ion{O}{1}]6364~\AA~line flux and multiply it by the branching ratio of 3.0 to obtain the line flux that is used to calculate H$_2$O production rates.  This was not required for our November observations, as the Doppler shift had decreased (in absolute value) so that the  [\ion{O}{1}]6300~\AA~line was no longer located on an O$_2$ telluric absorption (see Fig.~\ref{raw}), and spectra obtained on these dates have red line ratios consistent with the known branching ratio of 3.0.\\

\subsection{Spectral Fitting Model for OH}
\indent Unlike [\ion{O}{1}] emission, the OH emission exhibits ro-vibrational structure superimposed on the electronic transition, which results in an intricate band structure for the emission features (see Fig.~\ref{spectra}).  To fit the OH emission we employ the same modeling technique used by~\cite{McKay2016}, which we briefly describe here.  The program references a line list for the species of interest (in this case the OH A-X (0-0) band).  Once this line list is compiled, we fit each spectral feature in the list to a Gaussian profile.  We add all the line profile fits together to create an empirical fit to the spectrum and then integrate over this model fit to obtain a flux for the OH band.  While in general the program handles multiple species simultaneously, the OH A-X (0-0) band does not suffer from contamination of spectral features due to other species, therefore their inclusion is not necessary.  More details on the general program can be found in~\cite{McKay2014}, with the specific application to OH detailed in~\cite{McKay2016}.\\

\subsection{Conversion of Observed Flux to Production Rate}
We derived H$_2$O production rates based on two different spectral features: [\ion{O}{1}]6300, 6364~\AA~emission and the OH A-X (0-0) band at 3080~\AA.  For [\ion{O}{1}], we employ the same Haser model for the [\ion{O}{1}] emission used in~\cite{McKay2012} to infer H$_2$O production rates from our [\ion{O}{1}] observations.  For OH, we employ the same Haser model as in~\cite{McKay2016}, which has modifications that emulate the vectorial model (see~\cite{McKay2014} for more details), and adopt fluorescence efficiencies from~\cite{SchleicherAHearn1988} that account for the Swings effect (the dependence of the fluorescence efficiency on the heliocentric velocity of the comet).  The OH production rates are converted to H$_2$O production rates using the relation from~\cite{CochranSchleicher1993}:
\begin{equation}
Q_{H_2O}=1.361R_h^{-0.5}Q_{OH}
\end{equation}
where R$_h$ is heliocentric distance in AU.  The relation depends on heliocentric distance because the expansion velocity of H$_2$O in the coma is dependent on heliocentric distance while the ejection velocity of OH into the coma is not.  Parameters for the Haser models are given in Table~\ref{Haser}.\\

\section{Results}
\indent The resulting average H$_2$O production rates for each observation date are given in Table~\ref{AveQrates} and plotted in Fig.~\ref{Qrates}, along with other measured H$_2$O production rates for ISON from the literature.  In Table~\ref{AveQrates} we also present the surface area needed to provide the observed H$_2$O production (i.e. the active area).  We employ the sublimation model of~\cite{CowanAHearn1979} with a visual albedo of 0.05 and thermal emissivity of unity and the slow-rotator approximation.  The slow-rotator approximation is appropriate for objects with slow rotation rates or low thermal inertia, and is typically the model used to describe cometary nuclei~\citep[e.g.][]{Bodewits2014}.  Table~\ref{AveQrates} also displays the active fraction of the nucleus based on our active area calculations and assuming a spherical nucleus with a radius of 0.75 $\pm$ 0.15 km as derived by~\cite{Lamy2014} from HST observations (although they quote a more precise value of 0.68 $\pm$ 0.02 km,~\cite{Lamy2014} also note that this precise number is model dependent, so for this work we will employ their more conservative value of 0.75 $\pm$ 0.15 km).  Size estimates for Oort Cloud comets like ISON are very rare, so the size estimate for ISON provides a unique opportunity to constrain the active fraction of an Oort Cloud comet.  As both active area and active fraction are model dependent, Table~\ref{activearealowA} shows active areas and active fractions derived for isothermal, subsolar, and fast-rotator models in addition to the slow-rotator model to examine the effect of model assumptions on our results.  Table~\ref{activeareahighA} is the same as Table~\ref{activearealowA}, except for a visual albedo of 0.5 instead of 0.05.  The active area results are also plotted in Fig.~\ref{activearea}.  The derived active areas are as high as 100\% and in some cases even higher.  Implications for this will be discussed in the next section.\\

\section{Discussion}
\subsection{Evolution of H$_2$O Production}
\indent Our H$_2$O production rates show a very slow increase from 1.8 to 0.9 AU, with a dramatic rise in H$_2$O production at around 0.6 AU, in agreement with previous measurements as shown in Fig.~\ref{Qrates}.  Our observation at a heliocentric distance of 0.6 AU was only $\sim$ 12 hours after the first of several outbursts that were observed at heliocentric distances less than 0.6 AU~\citep[see][for an analysis of comet ISON's lightcurve and outbursts]{SekaninaKracht2014}.  The difference in H$_2$O production before and after the outburst is about a factor of 15, which matches well the observed increase in brightness during the outburst~\citep{SekaninaKracht2014}, suggesting that the outburst was indeed correlated to a massive increase in H$_2$O production.  We will discuss what led to this strong increase in gas production in the following sections.  The H$_2$O production rate seems to have leveled out between 0.6 and 0.44 AU, however we note that clouds may have influenced our derived fluxes on this date.  Although our derived production rate is consistent with other measurements around this time, the uncertainty introduced by clouds means the uncertainty in this measurement could be a factor of two or maybe even higher, making it difficult to derive any strong conclusions on the behavior of the H$_2$O production between 0.6 and 0.44 AU.

\subsection{Extended Sources of H$_2$O}
\indent Past observations of comets such as C/2009 P1 (Garradd) have shown trends in derived H$_2$O production rate with the projected field of view at the comet, with larger fields of view showing larger production rates.  This has been cited as evidence for an icy grain halo with an extent of 10$^4$-10$^5$ km that serves as an extended source of H$_2$O production~\citep{Combi2013, Bodewits2014, McKay2015}.  The Lyman-$\alpha$ observations of~\cite{Combi2014b} have very large fields of view ($\sim$ 10$^6$-10$^7$ km), while slit-based spectroscopy like the observations presented here and IR observations such as~\cite{DelloRusso2016} and~\cite{DiSanti2016} have much smaller projected fields (a few $\times$ 10$^3$ km), and narrowband OH imaging~\citep{KnightSchleicher2015, Opitom2014} has intermediate fields of view on the order of tens of thousands of km.  However, throughout the apparition for ISON there appears to be no correlation between projected field of view and derived H$_2$O production, with the possible exception of our [\ion{O}{1}] observations at R$_h$=1.8 AU and 1.6 AU and nearly concurrent measurements of OH from~\cite{KnightSchleicher2015}, which employ a field of view on the order of 5$\times$10$^4$ km (see Fig.~\ref{Qrates}).~\cite{KnightSchleicher2015} argue from analysis of the dust spatial profiles that before late October (i.e. at the time of our observations at R$_h$=1.8 AU and 1.6 AU) the coma did contain icy grains, so it is possible that there was an extended source of H$_2$O production at larger heliocentric distance that dissipated with the disappearance of the icy grains in late October.~~\cite{Combi2014b} reached a similar conclusion about the presence of an extended source of H$_2$O production in comet ISON from late October onward.  It is also possible that the extended source did not disappear, but rather became much smaller in spatial extent and was no longer spatially resolved by observations.  With derived H$_2$O production rates being consistent across platforms, any extended source present at R$_h$ $<$ 1.3 AU must have had a spatial extent smaller than the projected field of view of our observations ($<$ 1000 km).

\subsection{Analysis of the Active Area: Evidence for Fragmentation} 
\subsubsection{Evolution of the Active Area} 
\indent Our derived active areas (Table~\ref{AveQrates}) show a relatively constant value of ~$\sim$10 km$^2$ outside of R$_h$=1.2 AU, which then decreased to about half this value over the R$_h$=0.9-1.2 AU range, in agreement with calculations by~\cite{Combi2014b}, although~\cite{Combi2014b} have a slightly later onset ($\sim$1 AU) for their tentative detection of a decrease in active area.  For the nucleus radius of 0.75 $\pm$ 0.15 km determined by~\cite{Lamy2014}, the active areas derived for R$_h$ $>$ 0.9 AU are consistent with a very high active fraction for the nucleus of 50\% or even a completely active surface, though the error bars are quite large due to uncertainty in the actual size of the nucleus coupled with uncertainties in the H$_2$O production rate.  In addition, our calculations of the active fraction for ISON assume a spherical nucleus, which is not necessarily a valid assumption.  Although a precise measurement of the active fraction for ISON is difficult, both our results and those of~\cite{Combi2014b} suggest that it is 50-100\% before the first outburst at R$_h \sim$ 0.6 AU.  However, at R$_h$=0.6 AU, both we and~\cite{Combi2014b} find an active area of $\sim$ 30 km$^2$, much larger than the total surface area of a spherical nucleus with radius $\sim$ 0.75 km.  This implies that at this time there was a large fragmentation event, or at least shedding of a large amount of ice-rich material, that resulted in more surface area being exposed directly to sunlight.  There is also the possibility that the active area decreased to half this value by 0.44 AU, but as stated in Section 4.1 the water production rate derived on this date may be affected by clouds, making it difficult to arrive at any definitive conclusions about the active area at this time.  Comparison to other measurements of the water production rates in Fig. 2 and the active areas they imply suggest that it seems unlikely that the active area was higher at 0.44 AU than at 0.6 AU.

\subsubsection{Alternate Models}
\indent To test the robustness of this conclusion, we examine not only active areas from the slow-rotator model, but also isothermal, subsolar, and fast-rotator models.  The slow-rotator model is valid for slow rotation periods and/or low thermal inertia of the surface material.  These conditions mean that every point on the cometary surface is in thermal equilibrium with the solar radiation incident upon it.  Comet rotation periods vary anywhere from hours to days, with a possible value of 10.4 hours determined for ISON~\citep{Lamy2014}.  This does not fulfill the slow-rotator approximation.  However, the thermal inertias for comets where such measurements are available are quite low~\citep[e.g.][]{Groussin2013, Choukroun2015}, hence the tendency in the literature is to assume cometary nuclei follow the slow-rotator approximation.  Therefore the slow-rotator is our preferred model.  The isothermal model assumes the whole nucleus surface is the same temperature, while the subsolar model assumes the whole nucleus surface has the temperature of the subsolar point.  Although these models are idealizations that likely do not occur in reality, they provide bounds on the sublimation rate coming off the surface.  The fast-rotator model assumes that lines of latitude on the cometary nucleus are isotherms, which occurs for fast rotation periods and/or high thermal inertias of the surface material.

\indent The active areas and active fractions derived for the different models with a visual albedo of 0.05 are shown in Table~\ref{activearealowA} and Fig.~\ref{activearea}.  At the heliocentric distances of our observations ($<$ 2 AU, well within the H$_2$O sublimation line, such that H$_2$O can be assumed to be fully activated), the fast-rotator and isothermal approximations give very similar active areas, which are slightly higher than those derived from the slow-rotator model.  This discrepancy is maximum at the largest heliocentric distances, and decreases as the comet moves towards the Sun.  The subsolar point model produces significantly smaller values for the active area than the other models, and is the only model with active fractions less than 50\%.\\

\indent Because all of these models suggest ISON's active fraction was much higher than the typical 5\% seen in other comets~\citep[e.g.][]{AHearn1995}, this either suggests that a significant fraction of the comet's surface was active, or there was an extended source of H$_2$O in the coma such as sublimation from icy grains.  We discussed the possibility of an extended source in the previous section.  For the following we will explore the possibility of a surface that was 20-100\% active (encompassing values from all the thermal models).  This high an active fraction could suggest that a large fraction of the cometary surface was exposed H$_2$O ice, meaning assuming a visual albedo of 0.05 may not be a reasonable assumption.  Therefore we explored thermal models for the nucleus where both the visual albedo and thermal albedo of the nucleus have a much higher value of 0.50.  The results of these models are presented in Table~\ref{activeareahighA} and Fig.~\ref{activearea}.  Assuming a higher albedo shifts the derived active areas upward by about a factor of two compared to the lower albedo models.  This only strengthens the conclusion that ISON's active fraction was quite large and perhaps an icy grain source of H$_2$O is needed to account for the comet being hyperactive, though this source would have to have a small spatial extent (see section 4.2).\\

\indent We will now examine which of our thermal models is most likely to describe ISON's activity.  The subsolar model is the only model with active fractions less than 50\%, meaning this is the only model that does not require a nearly uniformly active surface or an extended source.  The subsolar model is only valid if all the activity is dominated by the subsolar point, which is generally not observed in comets.  A special case that would fit this description is where a polar source region is the dominant source of activity and the rotation pole is pointed directly at the Sun (i.e. the rotation pole is at the subsolar point).  Early in the apparition (Spring 2013, R$_h$=4.0 AU) there was evidence for a polar source region and a rotation pole pointed at the Sun based on HST~\citep{Li2013} and ground-based observations~\citep{KnightSchleicher2015} of a dust jet pointed in the solar direction, the  orientation of which did not change over time.  However, this feature dissipated as ISON moved towards the Sun and the CN jet morphology during October-November did not match well with a polar jet~\citep{KnightSchleicher2015}, making it likely that during the time of our observations the activity was not driven by a polar source region located at the subsolar point.  Therefore we do not find the subsolar model a likely description of ISON's activity.  We cannot rule out the plausibility of the other models with current results.\\

\subsubsection{Implications for ISON's Activity}
\indent The arguments of the previous section mean that ISON's surface was at least 50\% active or there was an extended source of H$_2$O production in the coma, most likely due to the presence of icy grains.  The high albedo (A=0.5) cases all favor an active area of more than 100\%, implying an extended source, in which case there is a degeneracy between the active area of the extended source and the active area of the nucleus.  The low albedo cases (A=0.05) are consistent with both a highly active nucleus or an extended source.  If an extended source is present it must have a small spatial extent (less than $\sim$1000 km, the projected field of view of our slit spectroscopy) as argued in section 4.2.  This is plausible, considering that 103P/Hartley 2 was shown to have a hyperactive nucleus and that this extra active area came from an extended source of icy grains in the coma~\citep{AHearn2011, Kelley2013}.  However, H$_2$O production rates derived from remote sensing observations agree over drastically different fields of view, from $\sim$500 km for slit spectroscopy~\citep{DelloRusso2011, Mumma2011, McKay2014} to 10$^4$-10$^5$ km for imaging of OH~\citep{KnightSchleicher2013} and 10$^6$ km for Lyman-$\alpha$~\citep{Combi2011}, suggesting that the spatial extent of the icy grain source was not much larger than the smallest fields of view used (in this case a few hundered km).  This is consistent with the expected lifetimes of micron-sized icy grains at the heliocentric distance of 103P/Hartley 2 at the time of observation (R$_h$ $\sim$1.1 AU) of only 100-1000 seconds~\citep{Beer2006}, which is comparable to the crossing time for icy grains moving at $\sim$1 km s$^{-1}$.  Therefore for ISON to exhibit a similar phenomenon at similar and smaller heliocentric distances to 103P/Hartley 2 is certainly plausible, although uncertainties in the active area and nucleus size preclude this from being demonstrated conclusively.  Using observations of the spatial distribution of H$_2$O and rotational temperature in the inner coma (nucleocentric distance $<$ 1000 km),~\cite{Bonev2014} argued that ISON was releasing icy grains at heliocentric distances less than 0.6 AU, but as they obtained no observations at larger heliocentric distances (in particular before the first outburst in mid-November), their observations do not place a constraint on an icy grain source before the outburst. 

\indent All models for ISON on November 15 display active fractions greater than unity (Fig.~\ref{activearea}, Tables~\ref{activearealowA} and~\ref{activeareahighA}), implying a large increase in the active area of the comet/extended coma source.  This could be due to either a catastrophic fragmentation of the nucleus or a large outburst ejecting icy grains into the coma.  As discussed above,~\cite{Bonev2014} suggested that after the first outburst at $R_h$ $\sim$ 0.6 AU sublimation from icy grains was a significant source of water production.  However, ~\cite{Steckloff2015} argued that the coma morphology during and after the outburst in mid-November is inconsistent with a simple explosive event such as amorphous-crystalline water ice transition or trapped hypervolatiles triggering the outbursts and ejecting icy grains into the coma.  They argued instead that ISON split into a swarm of fragments with a characteristic size of 100 m.  As they are all the same size, drifting between fragments would not show up in Earth-based observations of the coma morphology until days after the event, consistent with observations.  However, it is likely that both the fragmentation event and release of small (micron-sized) icy grains contributed to the increased water production. 

\indent Although the absolute values of the active area and active fraction are model dependent, the evolution of the active area over the course of the apparition is much less sensitive to the specific thermal model employed, as shown in Fig.~\ref{activearea}.  All models support a relatively constant active area during October (R$_h$ $>$ 1.2 AU), a decrease in late October and early November, and a massive increase in active area in mid-November.  This is very similar to the results presented in~\cite{Combi2014b}.  The massive increase in active area in mid-November was likely associated with the first of several fragmentation events~\citep{Steckloff2015}, which increased the sublimating surface area in the form of smaller fragments and may also have released micron-sized icy grains into the coma, further increasing the effective active area.

\indent It is also possible in mid-November that both thermal emission from sub-micron grains in the coma and reflected solar radiation from the dust coma could provide significant heating to the surface, meaning our simple thermodynamical models for the nucleus temperature and sublimation would not be accurate.  If this is the case, this would mean less active area is required to explain the observed H$_2$O production rates.  However, quantifying this effect requires detailed calculations that are beyond the scope of this work.

\subsection{Influence of O$_2$}
\indent Recently, the Rosetta mission discovered the presence of O$_2$ in the coma of 67P/Churyumov-Gerasimenko at an abundance of 4\% relative to H$_2$O~\citep{Bieler2015}.  This discovery was unexpected and means that O$_2$ photodissociation could be an unaccounted source of \ion{O}{1} in cometary comae.  Specifically of relevance to this work is that O$_2$ has a high branching ratio for releasing \ion{O}{1} in the $^1$D state upon photodissociation~\citep[0.85,][]{Huebner1992}, which will then radiatively decay and release [\ion{O}{1}]6300~\AA~line emission.  This could potentially influence attempts to use [\ion{O}{1}]6300~\AA~line emission as a proxy for H$_2$O production.  To test the potential influence of O$_2$ on [\ion{O}{1}]6300~\AA~line intensities we modified our Haser model for [\ion{O}{1}] to have O$_2$ be the parent of \ion{O}{1} instead of H$_2$O.  For this purpose we employed a branching ratio of 0.854 for photodissociation of O$_2$ $\rightarrow$ O($^3$P) + O($^1$D) and a photodestruction timescale of 2.08$\times$10$^5$ s at 1 AU~\citep{Huebner1992}.  We find that at the abundance of 4\% relative to H$_2$O observed by Rosetta at 67P by the ROSINA instrument~\citep{Bieler2015}, O$_2$ can only account for 5-10\% of the observed [\ion{O}{1}]6300 line flux.  As our uncertainties are 10-20\%, O$_2$ at an abundance of 4\% relative to H$_2$O does not significantly influence our derived H$_2$O production rates.  Additionally, the consistency between our production rates derived from [\ion{O}{1}] and OH, as well as agreement with other values already in the literature from methods other than [\ion{O}{1}], support the conclusion that any O$_2$ present in the coma of ISON did not influence our derived H$_2$O production rates.  Although this suggests [\ion{O}{1}] studies cannot be used to infer O$_2$ abundances from the ground (which is very challenging to do directly due to severe telluric absorption), it does show that [\ion{O}{1}]6300~\AA~line emission can still be used as a proxy for H$_2$O in comets for O$_2$/H$_2$O ratios similar to that measured by ROSINA.  This conclusion is valid for the small FOV employed for our observations, with the potential contribution of O$_2$ becoming more important for larger FOV.  Using Rosetta Alice UV stellar occultation data~\cite{Kenney2017} found higher O$_2$/H$_2$O ratios than those measured by ROSINA (some as high as 50\%).  For O$_2$/H$_2$O of 50\% we expect that O$_2$ photodissociation would be a significant contributor to the [\ion{O}{1}] emission, being responsible for approximately half of the emission.  It is possible that such a high O$_2$ abundance could artificially raise our derived H$_2$O production rates so that they agree with observations such as~\cite{Combi2014} with larger FOV's, and could hide the presence of an extended source.  However, the consistency between our OH and [\ion{O}{1}] derived H$_2$O production rates (which have similar FOV) argues against this scenario and therefore against such a high O$_2$ abundance for comet ISON.

\section{Conclusions}
\indent In this work we presented H$_2$O production rates for comet C/2012 S1 (ISON) derived from observations of [\ion{O}{1}] and OH emission during its inbound leg, covering a heliocentric distance range of 1.8-0.44 AU.  Our production rates are in agreement with previous measurements using a variety of different techniques.  As our [\ion{O}{1}]-derived H$_2$O production rates are consistent with other methods, we conclude that the contribution of O$_2$ photodissociation to the [\ion{O}{1}] emission is negligible compared to the contribution of H$_2$O (for abundances observed by the ROSINA instrument aboard Rosetta), meaning [\ion{O}{1}] emission can still be utilized as a reliable proxy for H$_2$O production in comets (though for the much higher abundances observed by Rosetta Alice the presence of O$_2$ could introduce systematic error).  The lack of any apparent dependence of derived production rate with instrument field of view suggests that any extended source of H$_2$O production must have had a spatial extent smaller than the smallest projected field of view observation (in this case $<$ 1,000 km).  We find that ISON had an evolving active area throughout the apparition, with a rapid increase at about R$_h$=0.6 AU corresponding to the first of three major outbursts ISON experienced inside of 1 AU, consistent with other studies.  This rapid increase in active area likely indicates a major fragmentation event of the nucleus.  The 5-10 km$^2$ active area observed outside of R$_h$=0.6 AU is consistent with a 50-100\% active fraction for the nucleus, much higher than typically observed for cometary nuclei.  Although the absolute value of the active area is somewhat dependent on the thermal model employed, the changes in observed active area  are consistent among models, and the conclusion of a 50-100+\% active fraction is robust for realistic thermal models of the nucleus.  However, the possibility of a spatially unresolved icy grain source cannot be discounted.

\ack
We are grateful to Lori Feaga and an anonymous reviewer for helpful comments that improved the quality of this manuscript. We thank John Barentine, Jurek Krzesinski, Chris Churchill, Pey Lian Lim, Paul Strycker, and Doug Hoffman for developing and optimizing the ARCES IRAF reduction script used to reduce these data.  We thank the APO and Keck observing staffs for their invaluable help in conducting the observations.  We are extremely grateful to Dr. Zlatan Tsvetanov for yielding some of his observing time so we could obtain the UT November 20 observations presented in this paper.  We thank Dr. Cyrielle Opitom for sharing some of her unpublished water production rates for inclusion in Fig.~\ref{Qrates}.  We would also like to acknowledge the JPL Horizons System, which was used to generate ephemerides for nonsidereal tracking of the comets during the observations, and the SIMBAD database, which was used for selection of reference stars.  This work was supported by the NASA Postdoctoral Program, administered by USRA, as well as the NASA Planetary Atmospheres Program, the NASA Planetary Astronomy Program, and the NASA Emerging Worlds Program.  The authors wish to recognize and acknowledge the very significant cultural role and reverence that the summit of Maunakea has always had within the indigenous Hawaiian community.  We are most fortunate to have the opportunity to conduct observations from this mountain.
\label{lastpage}


\bibliography{./references.bib}

\bibliographystyle{plainnat}

\clearpage

\begin{landscape}
\begin{table}
\begin{center}
\caption{
\label{observations}
\label{lasttable}
}
\textbf{Observation Log}\\
\begin{tabular}{lllllllllll}
\hline
Date (UT) & Instrument & $r$ (AU) & $\dot{r}$ (km s$^{-1}$) & $\Delta$ (AU) & $\dot{\Delta}$ (km s$^{-1}$) & FOV (km) & G2V & A0V & Flux Cal & Conditions\\
\hline
9/23/2013 & ARCES & 1.80 & -31.3 & 2.38 & -48.9 & 2760 $\times$ 5540 & HD30854 & HR3224 & HR 3454$^a$ & Clear\\
10/3/2013 & ARCES & 1.61 & -33.0 & 2.09 & -50.9 & 2430 $\times$ 4860 & HR 79078 & HR 3711 & HR 3454 & Clear\\
10/18/2013 & Tull Coude & 1.31 & -36.6 & 1.64 & -51.9 & 1430 $\times$ 9780 &  Solar Port & $\gamma$ Gem & $\gamma$ Gem & Clear\\
10/20/2013 & Tull Coude & 1.27 & -37.2 & 1.58 & -51.8 & 1380 $\times$ 9420 & Solar Port & $\gamma$ Gem & $\gamma$ Gem & Clear\\
10/21/2013 & Tull Coude & 1.25 & -37.5 & 1.55 & -51.7 & 1350 $\times$ 9240 & Solar Port & $\gamma$ Gem & $\gamma$ Gem & Clear\\
10/25/2013 & HIRESb & 1.15 & -39.1 & 1.42 & -50.8 & 890 $\times$ 7230 & Hyades 64 & HR 2207 & Hilt 600 & Clear\\
10/28/2013 & HIRESb & 1.08 & -40.3 & 1.33 & -49.8 & 830 $\times$ 6770 & Hyades 64 & HR 3134 & Hilt 600 & Clear\\
11/6/2013 & ARCES & 0.87 & -44.8 & 1.10 & -43.5 & 1280 $\times$ 2560 & HD 95868 & HR 4464 & HR 4468 & Clear\\
11/15/2013 & ARCES & 0.60 & -53.5 & 0.90 & -25.7 & 1050 $\times$ 2100 & HD 110747 & HR 4722 & HR 4963 & Clear\\
11/20/2013 & ARCES & 0.44 & -62.8 & 0.86 & -4.1 & 1000 $\times$ 2000 & HD 110747 & HR 4722 & HR 4963$^b$ & Passing Clouds\\
\hline
\end{tabular}
\end{center}
$a$ Flux Standard taken from UT October 3 observation\\
$b$ Flux Standard taken from UT November 15 observation
\end{table}
\end{landscape}

\clearpage

\begin{table}
\begin{center}
\caption{
\label{Haser}
}
\textbf{Parameter Values Used in the Haser Models}\\

\begin{tabular}{lcccc}
Molecule & $\tau_p$ (s)$^a$ & $\tau_d$ (s)$^a$ & $V_{ej}$ (km s$^{-1}$) & g-factor (ergs s$^{-1}$ molecule$^{-1}$)$^a$\\
\hline
OH & 1.3 $\times$ 10$^4$ & 2.1 $\times$ 10$^5$ & 1.02 & 3.5 $\times$ 10$^{-15}$\\
\ion{O}{1}$^b$ & 8.3 $\times$ 10$^4$ & - & - & -\\
\ion{O}{1}$^c$ & 1.3 $\times$ 10$^5$ & - & - & -\\ 
\hline
\end{tabular}
\end{center}
$a$ Given for $r$=1 AU.  The listed g-factor also accounts for the Swings effect.\\
$b$ For [\ion{O}{1}] from dissociation of H$_2$O into H$_2$ and O; branching ratio employed is 0.07~\citep{BhardwajRaghuram2012}\\
$c$ For [\ion{O}{1}] from dissociation of OH; branching ratio for H$_2$O to OH + H employed is 0.855~\citep{Huebner1992} and the branching ratio for OH to O + H is 0.094~\citep{BhardwajRaghuram2012}.\\ 
\end{table}

\clearpage
\begin{table}
\begin{center}
\caption{
\label{AveQrates}
}
\textbf{Production Rates}\\
\begin{tabular}{ccccc}
\hline
UT Date & R$_h$ (AU) & $Q_{H_2O}$ (10$^{27}$ mol s$^{-1}$) & Active Area (km$^2$)$^a$ & Active Fraction$^b$ (\%)\\
\hline
9/23/2013 & 1.80 & 7.65 $\pm$ 1.03 & 9.0 $\pm$ 1.2 & 127 $\pm$ 54 \\
10/3/2013 & 1.61 & 9.10 $\pm$ 1.15 & 7.9 $\pm$ 1.0 & 112 $\pm$ 47 \\
10/18/2013 & 1.31 &14.5 $\pm$ 3.5 & 7.5 $\pm$ 1.8 & 106 $\pm$ 49\\
10/20/2013 & 1.27 & 21.8 $\pm$ 4.9 & 10.5 $\pm$ 2.4 & 149 $\pm$ 68\\
10/21/2013 & 1.25 & 17.3 $\pm$ 3.9 & 8.0 $\pm$ 1.8 & 113 $\pm$ 52\\
10/25/2013 & 1.15 & 12.7 $\pm$ 2.4 & 4.8 $\pm$ 0.9 & 68 $\pm$ 30\\
10/28/2013 & 1.08 & 14.1 $\pm$ 2.4 & 4.7 $\pm$ 0.8 & 66 $\pm$ 29\\
11/6/2013 & 0.87 & 21.3 $\pm$ 4.3 & 4.3 $\pm$ 1.0 & 61 $\pm$ 28\\
11/15/2013 & 0.60 & 347 $\pm$ 29.1 & 31.8 $\pm$ 2.7 & 450 $\pm$ 180\\
11/20/2013 & 0.44 & 400 $\pm$ 29.0 & 19.2 $\pm$ 1.4 & 270 $\pm$ 110\\
\hline
\end{tabular}
\end{center}
$a$ Assuming an albedo of 0.05, thermal emissivity of unity, and a slow-rotator thermal model~\citep{CowanAHearn1979}.  See Tables~\ref{activearealowA} and~\ref{activeareahighA}  for different parameter values.
$b$ Assuming a spherical nucleus with radius of 0.75 $\pm$ 0.15 km~\citep{Lamy2014}.
\end{table}

\clearpage

\begin{landscape}
\begin{table}
\begin{center}
\caption{
\label{activearealowA}
}
\textbf{Active Area (km$^2$) and Active Fraction (\%) for 0.05 Albedo}\\
\begin{tabular}{ccccccccc}
\hline
Model & \multicolumn{2}{c}{Isothermal} & \multicolumn{2}{c}{Fast-Rotator} & \multicolumn{2}{c}{Slow-Rotator} & \multicolumn{2}{c}{Subsolar}\\
\hline
UT Date & Active Area & Active Fraction & Active Area & Active Fraction & Active Area & Active Fraction & Active Area & Active Fraction\\
\hline
9/22/2013 & 16.2 $\pm$ 2.2 & 230 $\pm$ 97 & 15.3 $\pm$ 2.1 & 217 $\pm$ 92 & 9.0 $\pm$ 1.2 & 127 $\pm$ 54 & 1.9 $\pm$ 0.3 & 27 $\pm$ 11\\
10/3/2013 & 12.2 $\pm$ 1.5 & 173 $\pm$ 73 & 11.8 $\pm$ 1.5 & 168 $\pm$ 70 & 7.9 $\pm$ 1.0 & 112 $\pm$ 47 & 1.7 $\pm$ 0.2 & 24 $\pm$ 10\\
10/18/2013 & 9.8 $\pm$ 2.4 & 139 $\pm$ 65 & 9.7 $\pm$ 2.3 & 137 $\pm$ 64 & 7.5 $\pm$ 1.8 & 106 $\pm$ 50 & 1.7 $\pm$ 0.4 & 24 $\pm$ 11\\
10/20/2013 & 13.5 $\pm$ 3.0 & 190 $\pm$ 87 & 13.3 $\pm$ 3.0 & 188 $\pm$ 86 & 10.5 $\pm$ 2.4 & 148 $\pm$ 68 & 2.4 $\pm$ 0.5 & 34 $\pm$ 16\\
10/21/2013 & 10.2 $\pm$ 2.3 & 144 $\pm$ 66 & 10.1 $\pm$ 2.3 & 143 $\pm$ 65 & 8.0 $\pm$ 1.8 & 113 $\pm$ 52 & 1.8 $\pm$ 0.4 & 26 $\pm$ 12\\
10/25/2013 & 5.9 $\pm$ 1.1 & 84 $\pm$ 37 & 5.9 $\pm$ 1.1 & 83 $\pm$ 37 & 4.8 $\pm$ 0.9 & 69 $\pm$ 30 & 1.1 $\pm$ 0.2 & 16 $\pm$ 7\\
10/28/2013 & 5.5 $\pm$ 0.9 & 78 $\pm$ 34 & 5.5 $\pm$ 0.9 & 78 $\pm$ 34 & 4.7 $\pm$ 0.8 & 66 $\pm$ 29 & 1.1 $\pm$ 0.2 & 15 $\pm$ 7\\
11/6/2013 & 4.8 $\pm$ 1.0 & 68 $\pm$ 31 & 4.8 $\pm$ 1.0 & 68 $\pm$ 31 & 4.3 $\pm$ 0.9 & 61 $\pm$ 27 & 1.0 $\pm$ 0.2 & 15 $\pm$ 7\\
11/15/2013 & 33.7 $\pm$ 2.8 & 480 $\pm$ 200 & 33.7 $\pm$ 2.8 & 480 $\pm$ 200 & 31.8 $\pm$ 2.7 & 450 $\pm$ 180 & 7.7 $\pm$ 60 & 110 $\pm$ 40\\
11/20/2013 & 19.8 $\pm$ 1.4 & 280 $\pm$ 110 & 19.9 $\pm$ 1.4 & 280 $\pm$ 110 & 19.2 $\pm$ 1.4 & 270 $\pm$ 110 & 4.7 $\pm$ 30 & 67 $\pm$ 27\\
\hline
\end{tabular}
\end{center}
\end{table}
\end{landscape}

\clearpage

\begin{landscape}
\begin{table}
\begin{center}
\caption{
\label{activeareahighA}
\label{lasttable}
}
\textbf{Active Area (km$^2$) and Active Fraction (\%) for 0.50 Albedo}\\
\begin{tabular}{ccccccccc}
\hline
Model & \multicolumn{2}{c}{Isothermal} & \multicolumn{2}{c}{Fast-Rotator} & \multicolumn{2}{c}{Slow-Rotator} & \multicolumn{2}{c}{Subsolar}\\
\hline
UT Date & Active Area & Active Fraction & Active Area & Active Fraction & Active Area & Active Fraction & Active Area & Active Fraction\\
\hline
9/22/2013 & 26.3 $\pm$ 3.5 & 370 $\pm$ 160 & 25.2 $\pm$ 3.4 & 360 $\pm$ 150 & 16.2 $\pm$ 2.2 & 230 $\pm$ 100 & 3.5 $\pm$ 0.5 & 49 $\pm$ 21\\
10/3/2013 & 20.7 $\pm$ 2.6 & 290 $\pm$ 120 & 20.3 $\pm$ 2.6 & 290 $\pm$ 120 & 14.4 $\pm$ 1.8 & 200 $\pm$ 90 & 3.2 $\pm$ 0.4 & 45 $\pm$ 19\\
10/18/2013 & 17.4 $\pm$ 4.2 & 250 $\pm$ 120 & 17.2 $\pm$ 4.2 & 240 $\pm$ 110 & 13.9 $\pm$ 3.4 & 200 $\pm$ 90 & 3.2 $\pm$ 0.8 & 45 $\pm$ 21\\
10/20/2013 & 23.9 $\pm$ 5.4 & 340 $\pm$ 160 & 23.7 $\pm$ 5.3 & 340 $\pm$ 150 & 19.5 $\pm$ 4.4 & 280 $\pm$ 130 & 4.5 $\pm$ 1.0 & 63 $\pm$ 29\\
10/21/2013 & 18.2 $\pm$ 4.1 & 260 $\pm$ 120 & 18.0 $\pm$ 4.1 & 260 $\pm$ 120 & 14.9 $\pm$ 3.4 & 210 $\pm$ 100 & 3.4 $\pm$ 0.8 & 49 $\pm$ 22\\
10/25/2013 & 10.7 $\pm$ 2.0 & 150 $\pm$ 70 & 10.6 $\pm$ 2.0 & 150 $\pm$ 70 & 9.0 $\pm$ 1.7 & 130 $\pm$ 60 & 2.1 $\pm$ 0.4 & 30 $\pm$ 13\\
10/28/2013 & 10.1 $\pm$ 1.7 & 140 $\pm$ 60 & 10.0 $\pm$ 1.7 & 140 $\pm$ 60 & 8.7 $\pm$ 1.5 & 120 $\pm$ 50 & 2.0 $\pm$ 0.3 & 29 $\pm$ 13\\
11/6/2013 & 8.9 $\pm$ 1.8 & 130 $\pm$ 60 & 8.9 $\pm$ 1.8 & 130 $\pm$ 60 & 8.1 $\pm$ 1.6 & 120 $\pm$ 50 & 2.0 $\pm$ 0.4 & 28 $\pm$ 12\\
11/15/2013 & 63.1 $\pm$ 5.3 & 890 $\pm$ 370 & 63.1 $\pm$ 5.3 & 890 $\pm$ 370 & 60.0 $\pm$ 5.0 & 850 $\pm$ 350 & 14.7 $\pm$ 1.2 & 210 $\pm$ 90\\
11/20/2013 & 37.4 $\pm$ 2.7 & 530 $\pm$ 210 & 37.4 $\pm$ 2.7 & 530 $\pm$ 210 & 36.4 $\pm$ 2.6 & 510 $\pm$ 210 & 9.0 $\pm$ 0.7 & 130 $\pm$ 50\\
\hline
\end{tabular}
\end{center}
\end{table}
\end{landscape}

\clearpage

\begin{center}
 Figure Captions
\end{center}

\begin{figure}[p!]
\begin{center}
\includegraphics[width=\textwidth]{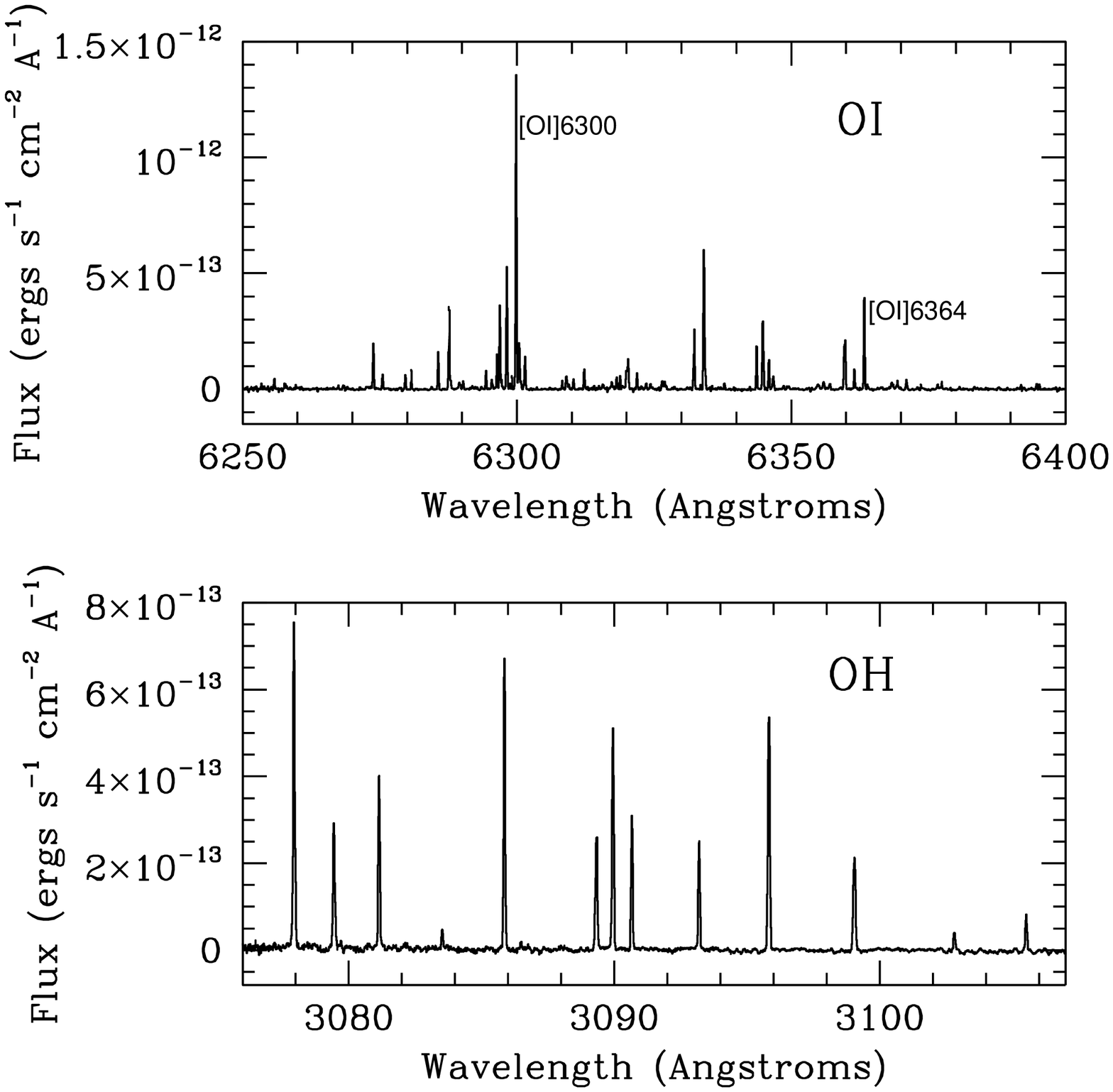}
\caption{
\label{spectra}
\label{lastfig}			
}
\end{center}
\end{figure}

Fig~\ref{spectra}: Top: Spectrum showing the [\ion{O}{1}] line region on November 15.  The [\ion{O}{1}]6300~\AA~and [\ion{O}{1}]6364~\AA~lines are labeled.  Most other emission features present are due to NH$_2$.  Bottom: Spectrum showing the OH A-X band on October 25.  Error bars have been omitted from both plots for clarity.

\begin{figure}[p!]
\begin{center}
\includegraphics[width=\textwidth]{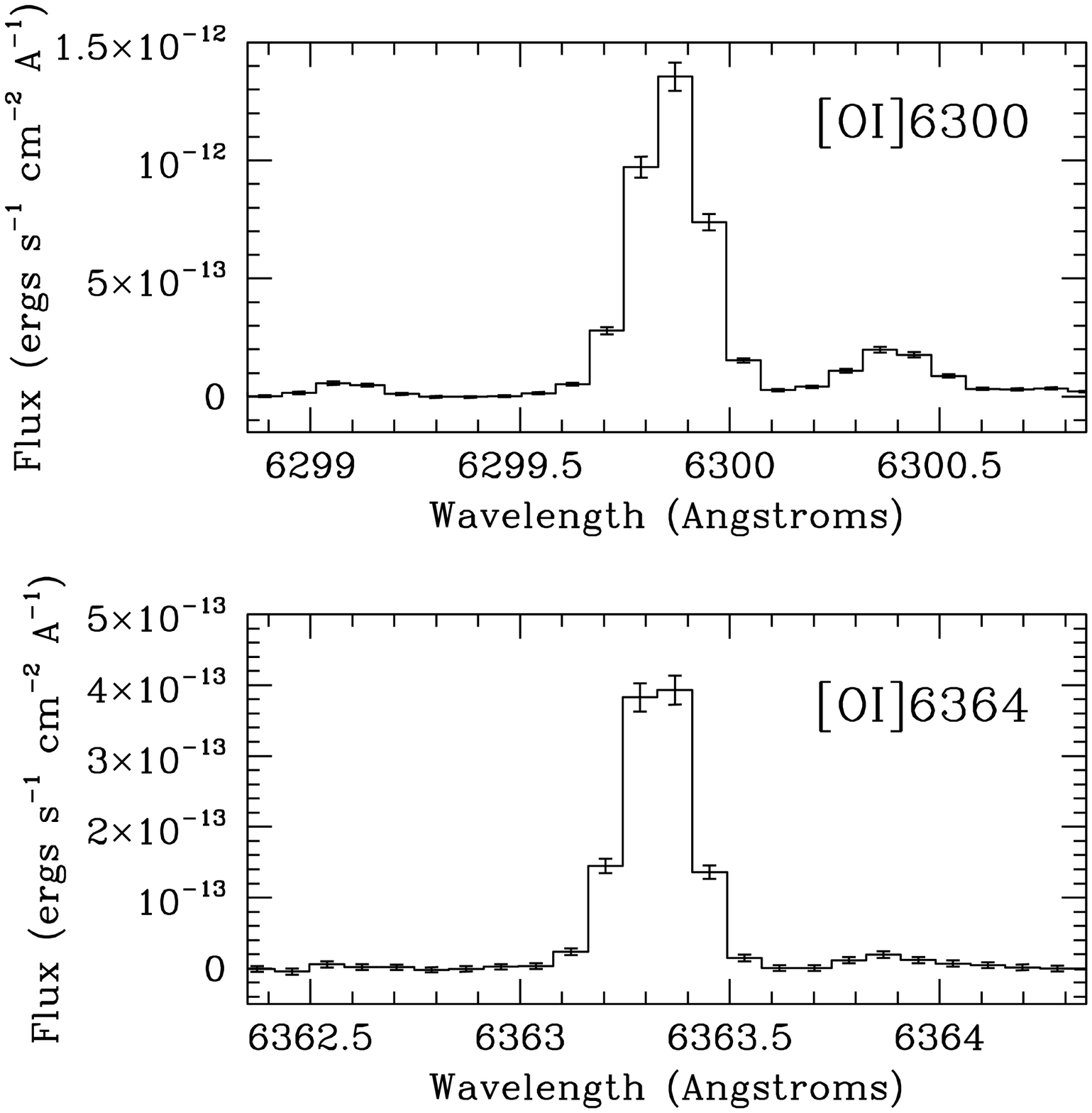}
\caption{
\label{zoom}
\label{lastfig}			
}
\end{center}
\end{figure}

Fig~\ref{zoom}: Top: Spectrum showing the [\ion{O}{1}]6300~\AA~line on November 15. The telluric line is the weaker feature redward of the cometary line.  Bottom: Spectrum showing the [\ion{O}{1}]6364~\AA~line on November 15. The telluric line is the weaker feature redward of the cometary line.

\begin{figure}[p!]
\begin{center}
\includegraphics[width=\textwidth]{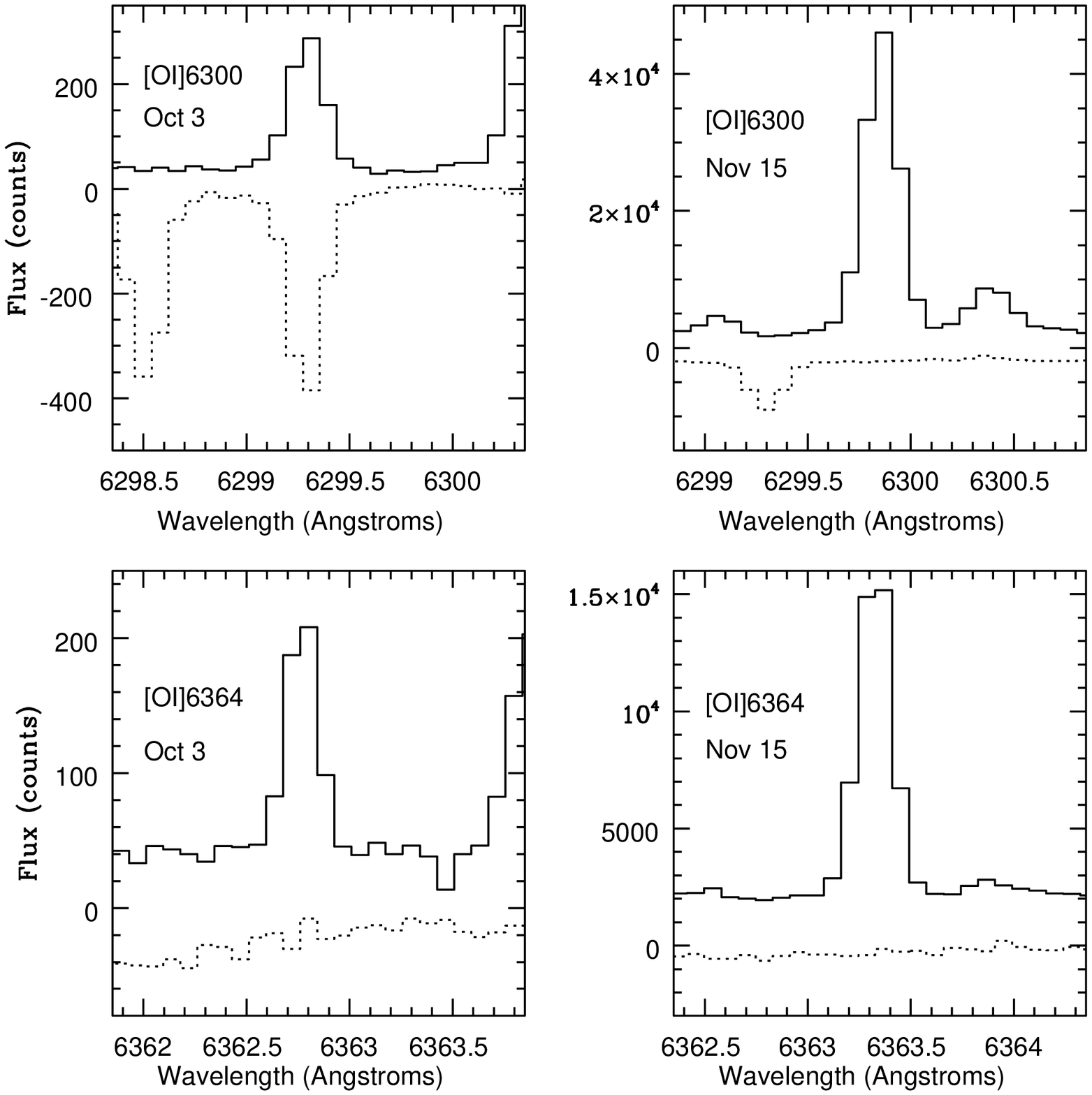}
\caption{
\label{raw}
\label{lastfig}			
}
\end{center}
\end{figure}

\begin{figure}[p!]
\begin{center}
\includegraphics[width=\textwidth]{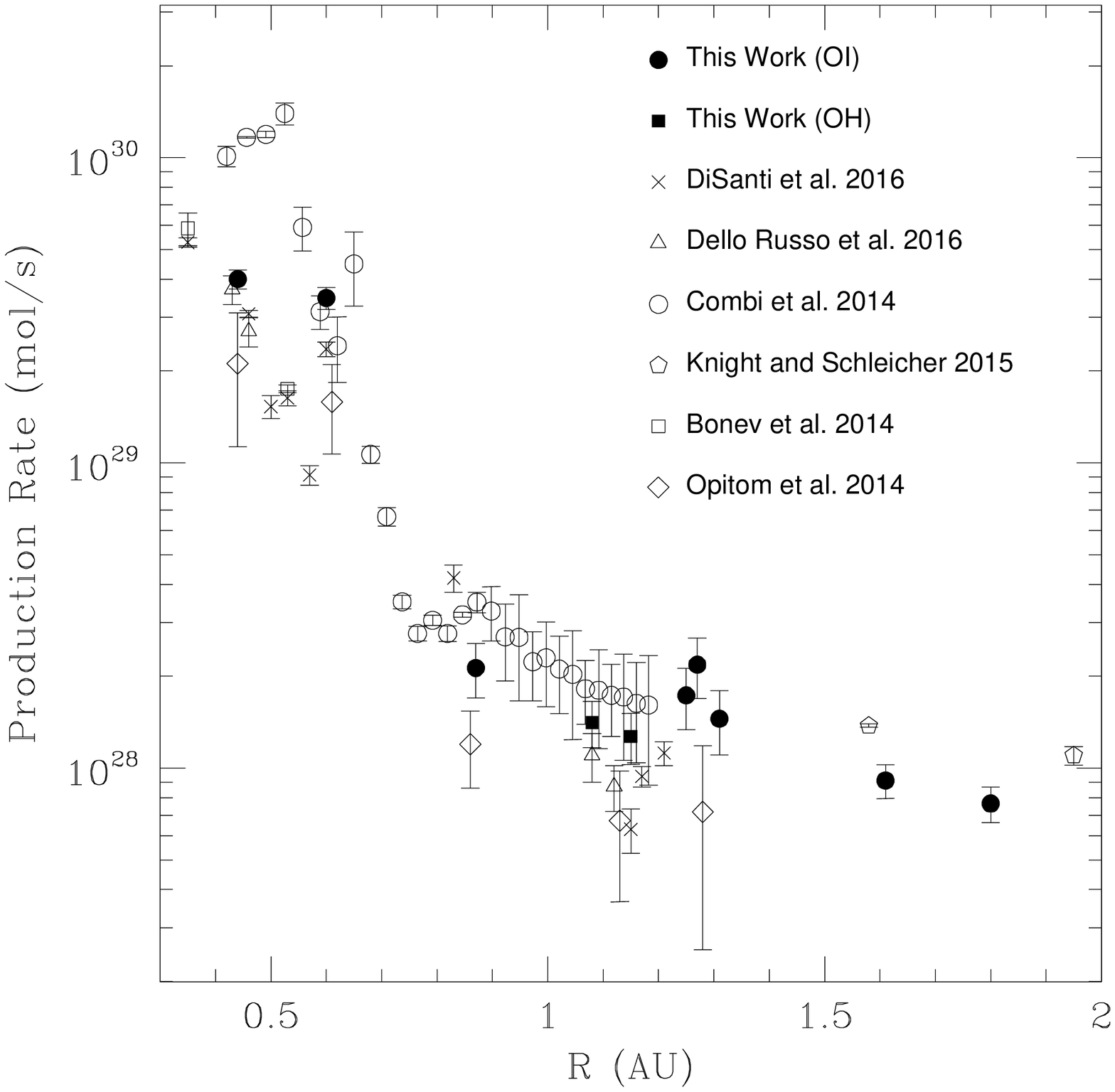}
\caption{
\label{Qrates}
\label{lastfig}			
}
\end{center}
\end{figure}

Fig~\ref{raw}: Raw extracted spectra (i.e. no telluric removal, solar subtraction, or flux calibration) showing the [\ion{O}{1}]6300~\AA~line (top row) and the [\ion{O}{1}]6364~\AA~line (bottom row) on UT October 3 (left column) and UT November 15 (right column), along with raw extracted spectra of the telluric standard on each date.  The telluric standard has been shifted down and for October 3 the counts were scaled by a factor of 10 to facilitate comparison with the comet spectrum.  On October 3 the [\ion{O}{1}]6300~\AA~line is directly coincident with an O$_2$ telluric absorption feature, while by November 15 the cometary line is no longer coincident with any O$_2$ telluric absorption.  The placement of the cometary line relative to the O$_2$ absorption is similar for all our September and October observations.  There are no strong telluric features underneath the [\ion{O}{1}]6364~\AA~line on either date.  Therefore for our September and October observations we calculate the H$_2$O production rate using the the [\ion{O}{1}]6364~\AA~line rather than the [\ion{O}{1}]6300~\AA~line.

Fig~\ref{Qrates}: Plot of H$_2$O production in comet ISON as a function of heliocentric distance.  Production rates from this work are plotted as filled symbols, with circles denoting production rates based on [\ion{O}{1}] observations and squares denoting production rates based on OH.  Other results from the literature are plotted for comparison as empty symbols.  Our values are in agreement with other measurements, and extend coverage of ISON's H$_2$O production rate to larger heliocentric distances than many previously published results. 

\begin{figure}[p!]
\begin{center}
\includegraphics[width=\textwidth]{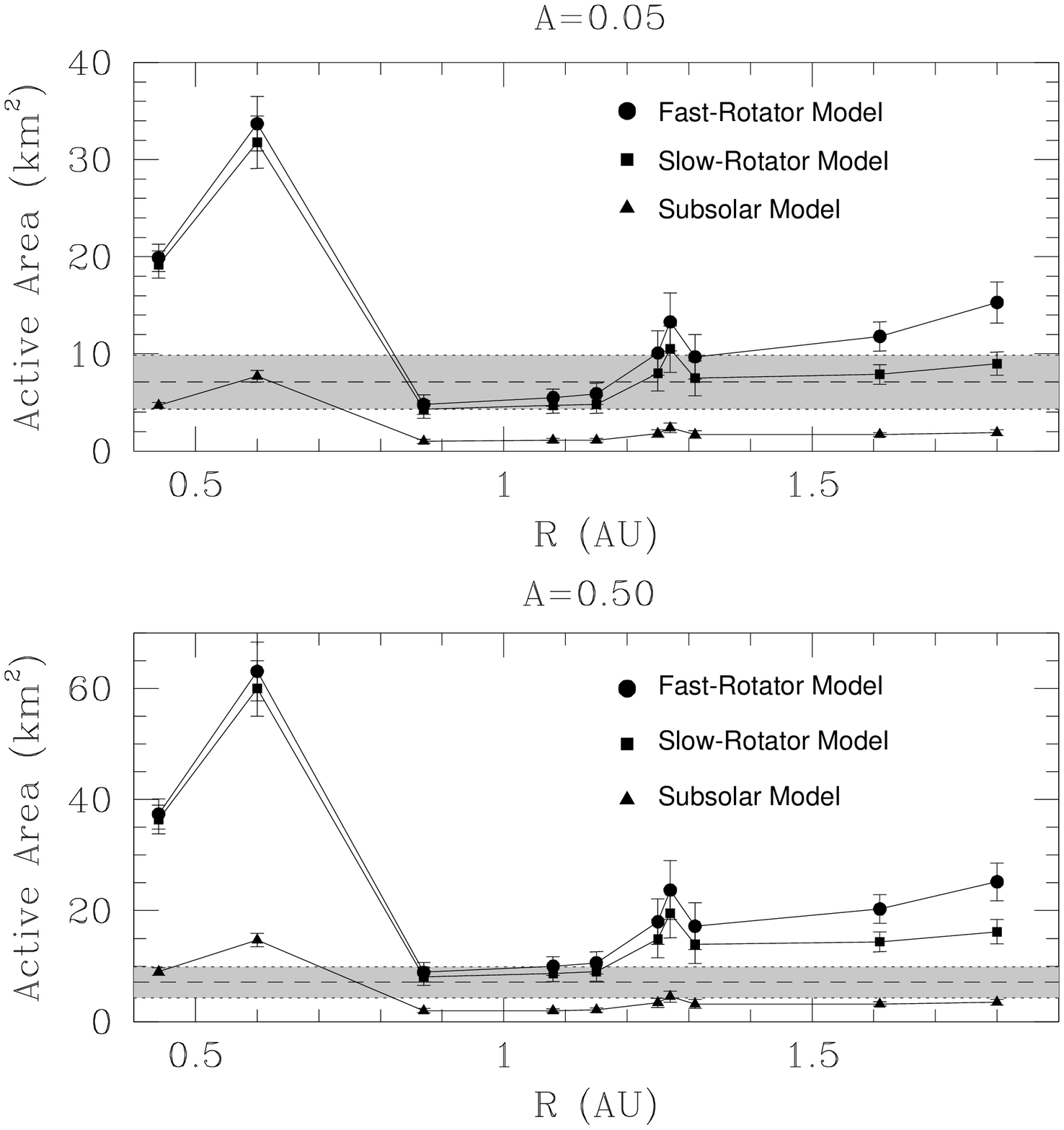}
\caption{
\label{activearea}
\label{lastfig}			
}
\end{center}
\end{figure}

Fig~\ref{activearea}: Derived active area as a function of heliocentric distance for different thermal models and albedos.  The top panel shows results for an albedo of 0.05, while the bottom panel shows results for a much larger albedo of 0.50.  For both albedos the isothermal and fast-rotator models give nearly identical values, so for clarity the isothermal model has not been plotted.  The dashed line is the total surface area of a spherical nucleus with a radius of 0.75 km~\citep{Lamy2014} and the shaded area depicts the uncertainty in this measurement.  Therefore active areas that fall in the shaded region are consistent with an active fraction of 100\%, while those above the shaded region indicate ISON is hyperactive, and additional surface area besides that of the nucleus is required to explain the observed H$_2$O production rates.  Changing the albedo and employing different thermal models changes the absolute value of the derived active area, but trends over the apparition are independent of the model employed.  The fast-rotator and slow-rotator models suggest an active fraction of at least 50\%, which could be indicative of a largely active surface or the presence of icy grains in the coma.  The subsolar model gives much lower active areas, but we believe this model does not provide a realistic description of ISON's activity (see Section 4.3.2).

\clearpage

\end{document}